\documentstyle[ppeuc,epsfig]{article}
\newcommand{\be}{\begin{equation}}
\newcommand{\ee}{\end{equation}}
\newcommand{\C}{{\boldmath C}}
\newcommand{\x}{{\boldmath x}}
\newcommand{\F}{{\boldmath F}}
\newcommand{\A}{{\boldmath A}}
\newcommand{\btheta}{{\boldmath \theta}}
\newcommand{\rgl}{\rangle}
\newcommand{\lgl}{\langle}

\begin{document}

\centerline{\small{
Published in Proceedings of the Particle Physics and Early Universe 
Conference (PPEUC), University of Cambridge, 7-11 April 1997.}}

\title{Cosmological Parameter Estimation from the CMB}
\author{Andy Taylor$^1$, Alan Heavens$^1$, Bill Ballinger$^1$, Max Tegmark$^2$}
\institute{$^1$Institute for Astronomy,
        University of Edinburgh, Royal Observatory,
        Blackford Hill, Edinburgh, EH9 3HJ, U.K., $^2$Institute for Advanced 
Study, Princeton, U.S.A.\\
ant@roe.ac.uk, afh@roe.ac.uk, web@roe.ac.uk, max@ias.edu}  

\begin{abstract}
We discuss the problems of applying Maximum Likelihood methods
to the CMB and how one can make it both efficient and optimal.
The solution is a generalised eigenvalue problem that allows
virtually no loss of information about the parameter being estimated, but 
can allow a substantial compression of the data set.  We discuss the 
more difficult question of simultaneous estimation of many parameters,
and propose solutions.   A much fuller account of most of this work is 
available (Tegmark et al. 1997, hereafter \cite{TTH}).

\end{abstract}

\section{Likelihood Analysis}

The standard method for extracting cosmological parameters 
from the CMB is through the use of Maximum Likelihood
methods. In general the likelihood function, ${\cal L}$,
for a set of parameters, $\btheta$, is given by a hypothesis, 
$H_x$, for the distribution function of the data set. In the case
of uniform prior, and assuming a multivariate Gaussian distributed data set
consistent with Inflationary models,
 the {\em a posteriori} probability distribution for the parameters is
\be
{\cal L} (\btheta | x, H_x) = (2 \pi)^{-n/2} |C(\btheta)|^{-1/2}
			\exp \left[- \frac{1}{2} \x^\dag \C(\btheta)^{-1} \x
\right],
\ee
where $\btheta=(Q,h,\Omega_0,\Omega_\Lambda,\Omega_b,...)$
are the usual cosmological parameters we would like to determine.
Examples of data are $\x=\Delta_T$ or $a_{\ell,m}$ and the statistics
of the $n$ data are fully parametrised by the data covariance matrix,
$\C(\btheta) = \lgl \x \x^\dag \rgl$.  For simplicity here we assume the
data have zero means.

\section{Problems with the likelihood method}

Two important questions we would like to settle about likelihood 
analysis are (a) is the method optimal in the sense that we
get the minimum variance (smallest error bars) for a given amount of data? 
and (b) is the method efficient -- can we realistically find the best-fitting 
parameters? As an 
example of this last point, if we have $n$ data points (pixels,
harmonic coefficients, etc), and $m$ parameters to estimate with a 
sampling rate of $1/q$, we find that the calculation time scales as
\be
	\tau \propto q^m \times n^3
\ee
where the first term is just the total number of points at which we need
to calculate the likelihood, and the second term is the time 
that it takes to calculate the inverse of $\C$ and its determinant.
Of course, in practice one would not find the maximum likelihood
solution this way, but it serves to illustrate the point.
Note that the covariance matrix depends on the parameters and therefore must be
evaluated locally in parameter space. For MAP or Planck we have 
$m \sim 11$, $q \sim 10$ and $n \sim 10^7$, resulting in $\tau
\sim H_0^{-1}$, even for nanosecond technology. But before we give up in 
dismay, it is worth looking a bit further at the theory of parameter 
estimation.
 
\section{Parameter information and the solution to our problems}

Suppose we have found the maximum likelihood solutions for each parameter,
$\btheta=\btheta_0$, then the likelihood function can be 
approximated by another multivariate Gaussian about this point;
\be
{\cal L}(\btheta|\btheta_0, H_\btheta) = (2 \pi)^{-m/2}
		|F|^{1/2} \exp\left[ - \frac{1}{2} \delta \btheta^\dag F
		\delta \btheta\right],
\ee
where $ \delta \btheta = \btheta - \btheta_0$ is the distance to the 
maximum in
parameter space and the parameter covariance matrix is given 
by the inverse of $\F$, the Fisher Information matrix;
\be
	F_{ij} = \lgl \delta \btheta_i \delta \btheta_j \rgl^{-1}
	= \frac{1}{2} {\rm Tr} [\A_i \A_j],
\ee
(if the means of the data are dependent on the parameters, this is 
modified -- see \cite{TTH}).  The far right hand side expression can be 
calculated for Gaussian
distributed data sets (ie equation 1), where 
$\A_i \equiv \partial \ln \C(\btheta)/\partial\theta_i$ is the slope of the log
of the data covariance matrix in parameter space.

By considering the Fisher matrix as the information content contained 
in the data set about each parameter, we see that the 
solution to our problem is to reduce the data set without changing
the parameter information content. Hence to
solve the problem of efficiency, we need to make a linear transformation
of the data set
\be
	\x' = B x,
\ee
where $B$ is a $n' \times n$ matrix where $n' \leq n$, and so $\x'$
may be a smaller data set than $\x$. If $n' < n$ the
transformation is not invertible and some information 
about the data has been lost. To ensure that the lost information
does not affect the parameter estimation (requirement (a)), we also require 
\be
	\frac{\partial F'}{\partial B} =0,
\ee
where $F'=BFB^\dag$ is the transformed Fisher matrix.
In order to avoid learning the unhelpful fact that no data is an optimal
solution, we add in the constraint that data exists. Since we have the
freedom to transform the data covariance matrix, we add the
constraint $\lambda(BCB^\dag-I)$, where $I$ is the unit matrix and 
$\lambda$ is a Lagrangian multiplier.

It can be shown (\cite{TTH}) that this is equivalent to a generalised
Karhunen-Lo\`eve eigenvalue problem, which has a unique solution
$B$ for each parameter. These solutions have the property that
\be
	B (\partial_{\theta_i} C) B^\dag= \lambda_i I,
\ee
where $\lambda_i=1/\sigma'_i$ are the eigenvalues
of the transformed data set and the inverse errors 
associated with each eigenmode of the new data set.

The new, compressed data set, $\x'$, can now be ordered by 
decreasing eigenvalue, so that the first eigenmode contains 
the most information about the desired parameter, the second
slightly less information, and so on. The total error on the
parameter is then simply given by the inverse of the $1\times1$ Fisher matrix 
\be
	F' = \frac{1}{2}\sum_{i=1}^{n'}
	 \left( \frac{1}{\sigma'^2_i} \right).
\ee

We are now free to choose how many eigenmodes to include in
the likelihood analysis. A 
compression of 10 will lead to a time saving of $10^3$.
However this is only exact if we know the true value of the parameters
used to calculate $B$. But if we are near the maximum likelihood solution
then we can iterate towards the exact solution.

This procedure is optimal for all parameters -- linear and nonlinear --
in the model. In the special case of linear parameters that are
just proportional to the signal part of the data covariance matrix
(for example the amplitude of $C_\ell$, if the data are the 
$a_{\ell m}$), the eigenmodes reduce to signal-to-noise
eigenmodes (\cite{bond}, \cite{bunn}). Hence our eigenmodes are more general than 
signal-to-noise eigenmodes. Furthermore, as our eigenmodes
satisfy the condition that the Fisher matrix is a maximum,
they are the optimal ones for data compression. Any other
choice, including signal-to-noise eigenmodes, would give a 
higher variance.

In Figure 1 we plot the uncertainty on 3 parameters for COBE-type
data, the quadrupole, $Q$, the spectral index of scalar 
perturbations, $n$ and the re-ionization optical depth, $\tau$.

\begin{figure}
 \htmlimage{scale=1.5,thumbnail=0.5}
 \centerline{\epsfig{file=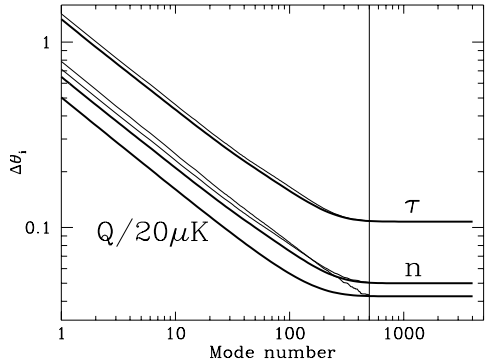,width=5cm,angle=270,clip=}}
 \caption{The 3 heavy lines show the error bars on 3 CMB parameters as
a function of the number of modes used.  Each set of modes has been optimised
for the parameter in question.  Note that approximately 400 modes are
all that is required to get virtually all the information from the entire
4016 cut COBE dataset.  The thin lines show the conditional errors 
from the SVD 
procedure outlined in section 5:  virtually all the (conditional) information 
on {\em all 3} parameters is obtained from the best 500 SVD modes.}
\end{figure}

\section{Estimating many parameters at once}

The analysis presented so far is strictly optimal only for
the conditional likelihood -- the estimation of one parameter when all
others are known.  A far more challenging task is to optimise the
data compression when all parameters are to be estimated from the 
data.  In this case, the marginal error on a single parameter $\theta_i$ rises
above the conditional error $1/\sqrt{F_{ii}}$ to $\sqrt{F_{ii}^{-1}}$.
As far as we are aware, there is no general solution known to this problem,
but here we present some methods which have intuitive motivation and
appear successful in practice.

Suppose that we repeat the optimisation procedure, outlined above, $m$ times,
once for each parameter.  The union of these sets should do well at estimating
all parameters, but the size may be large.  However, many of the modes
may contain similar information, and this dataset may be trimmed further
without significant loss of information.  This is effected by a singular value
decomposition of the union of the modes, and modes corresponding to small
singular values are excluded.   Full details are given in \cite{TTH}, 
and an example 
from COBE is illustrated in Figure 1, which 
shows that for the {\em conditional} likelihoods at least, the 
data compression procedure can work extremely well.  However, this on its
own may not be sufficient to achieve small marginal errors, especially
if two or more parameters are highly correlated.   This is expected to be the 
case for high-resolution CMB experiments such as MAP and Planck (e.g. for
parameters $Q_{\rm rms}$ and $n$).  To give a more concrete example -- 
a thin ridge of likelihood at $45^\circ$ to two parameter axes has small 
conditional errors, but the marginal errors can be very large.  This 
applies whether or not the likelihood surface can be approximated well
by a bivariate Gaussian.

\begin{figure}[h]
 \htmlimage{scale=1.5,thumbnail=0.5}
 \centerline{\epsfig{file=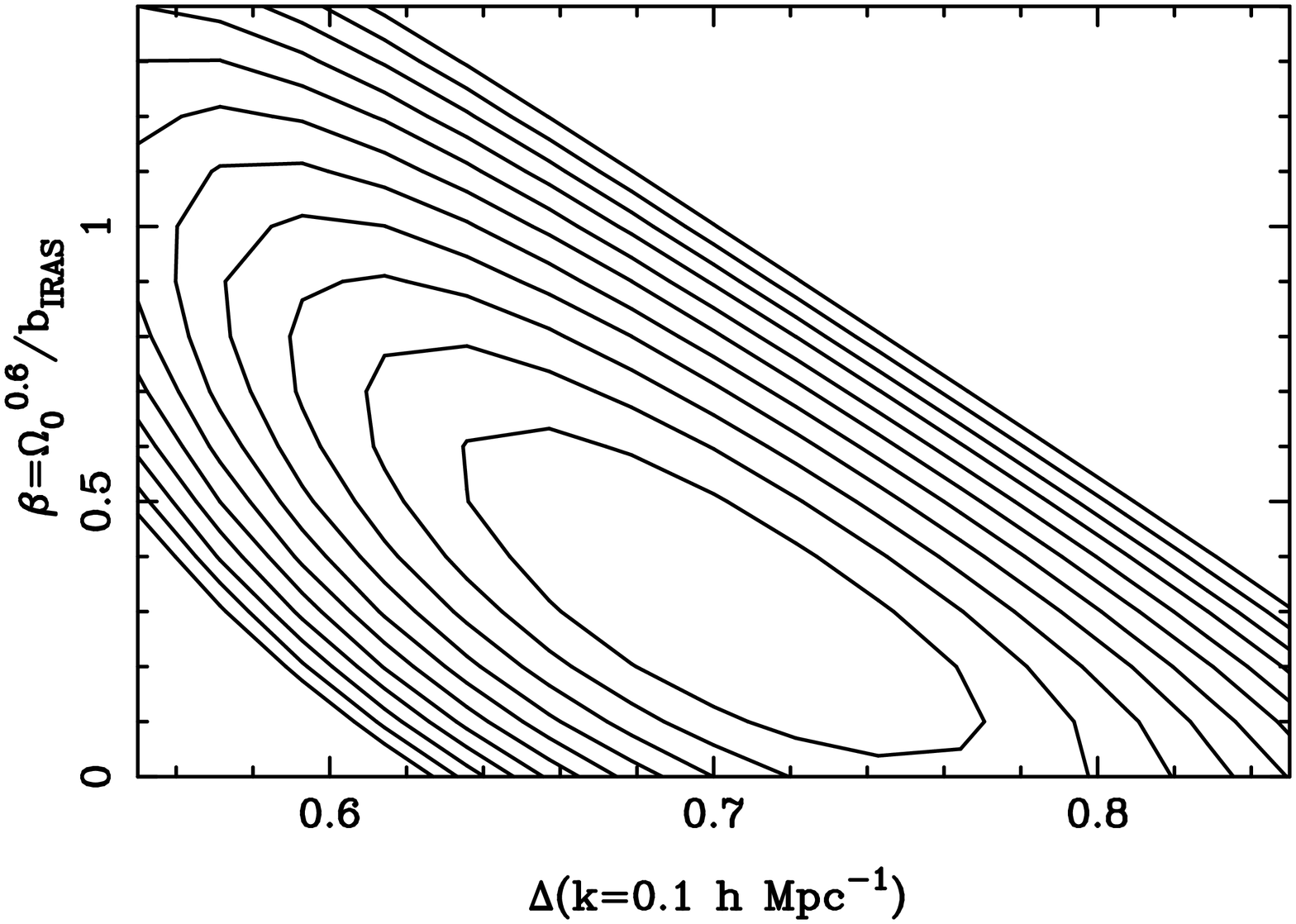,width=5cm,angle=270,clip=}
\epsfig{file=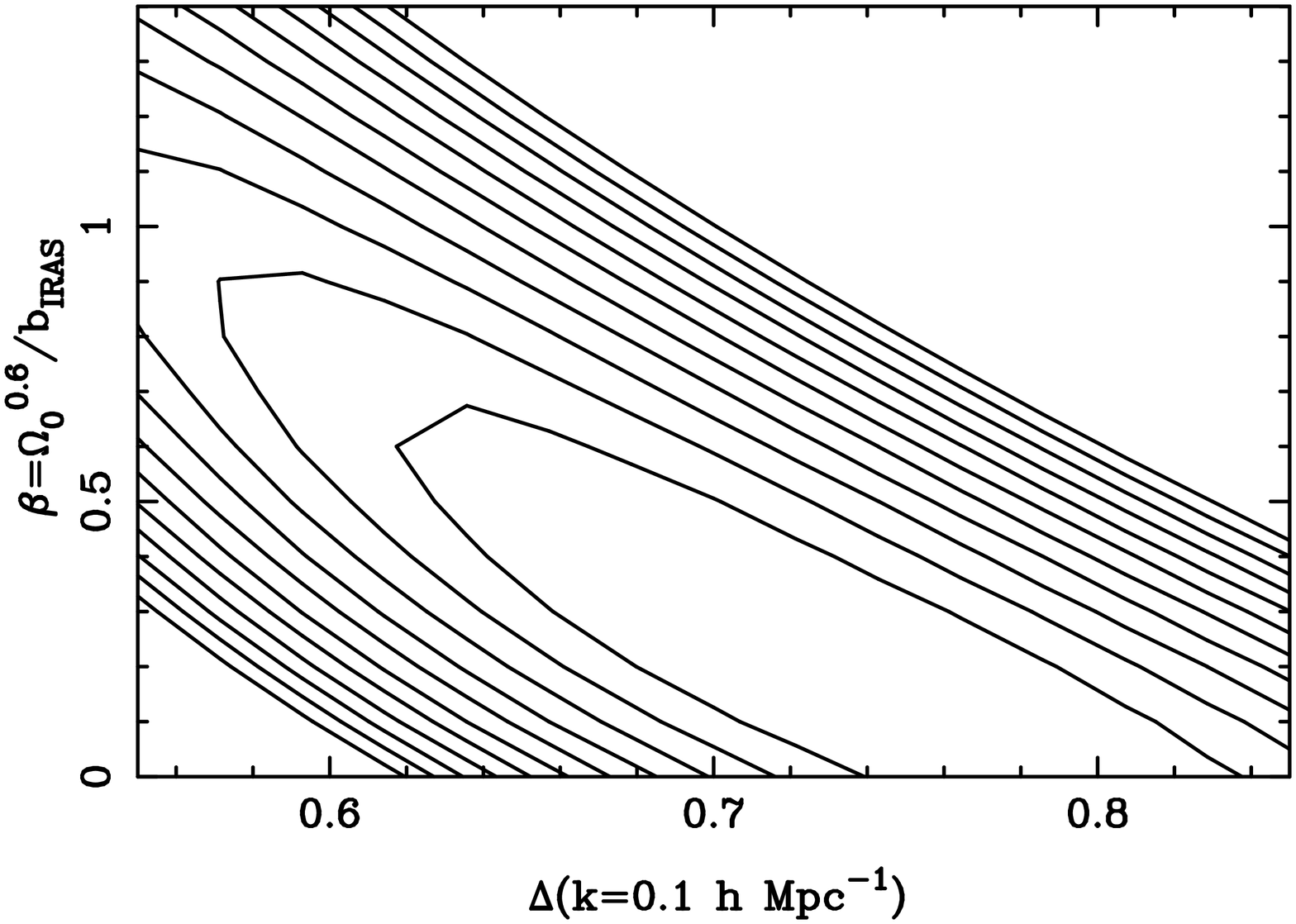,width=5cm,angle=270,clip=}}
\centerline{
\epsfig{file=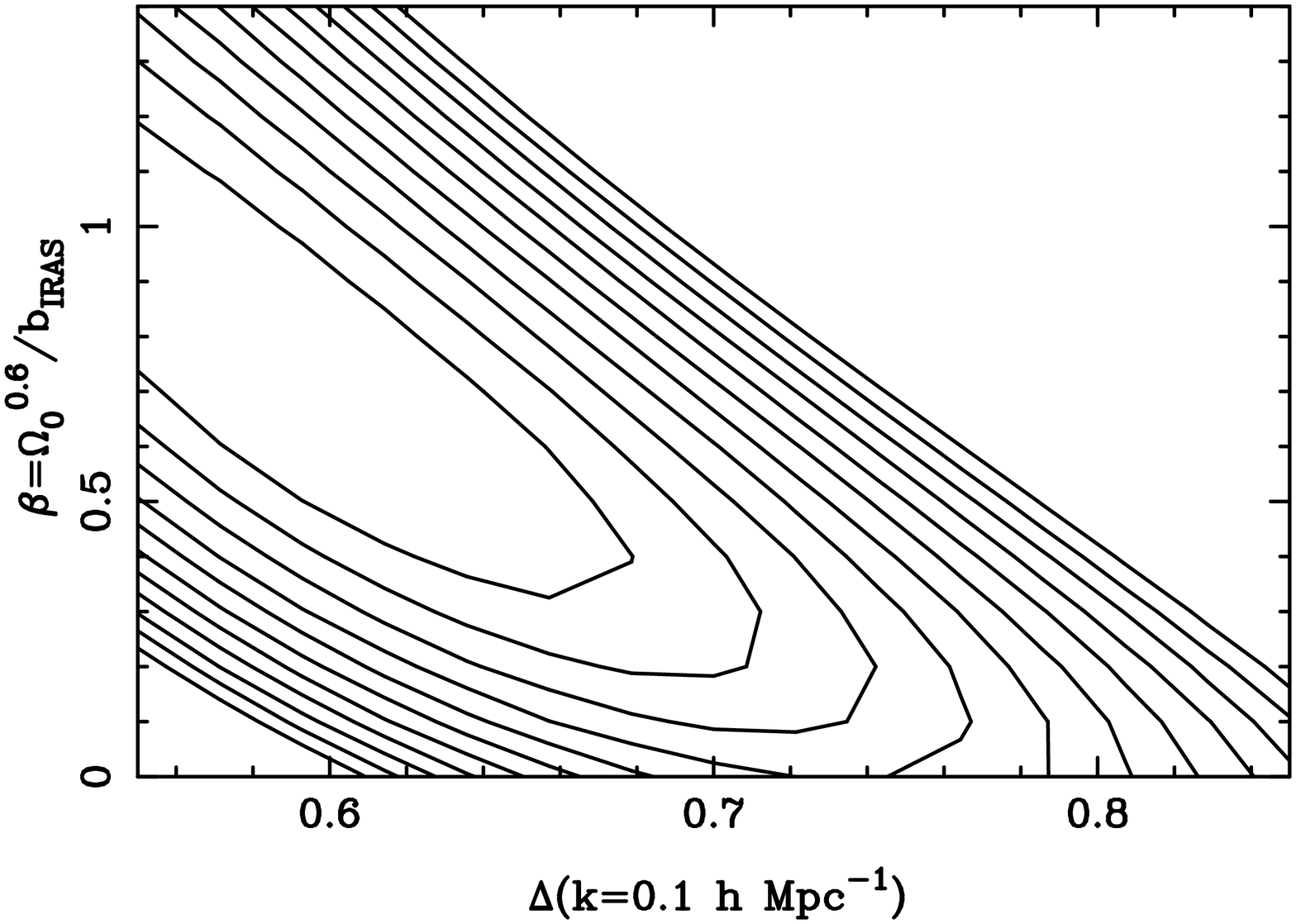,width=5cm,angle=270,clip=}
\epsfig{file=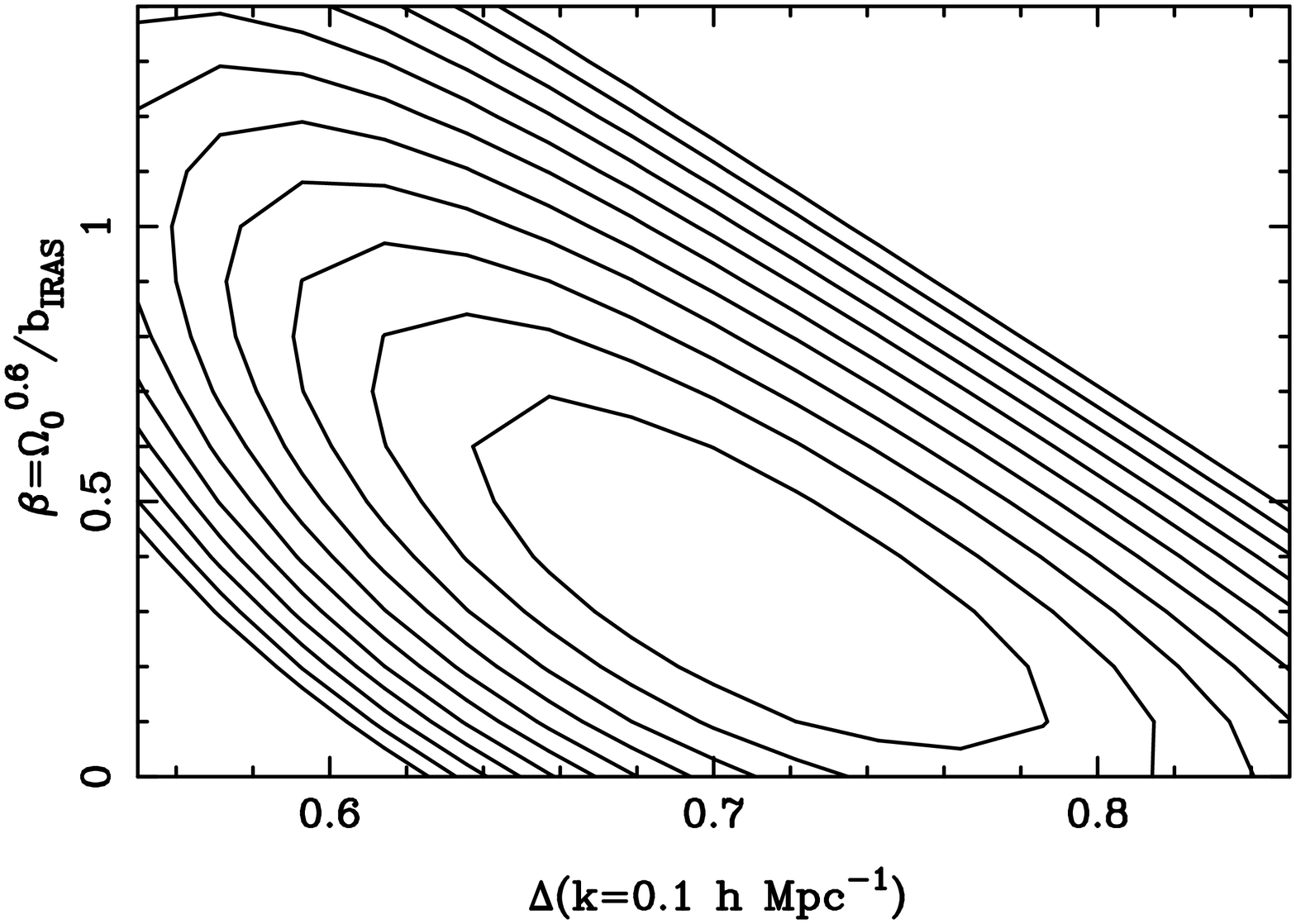,width=5cm,angle=270,clip=}
}
 \caption{Illustration of data compression with different algorithms.  
Top left: `Full' dataset of 508 modes (for details of parameters etc, see 
text).  Top right:  Best 320 modes 
optimised for measuring $\beta$.  Bottom left: Best 320 modes from
SVD application to modes optimised for $\beta$ and $\Delta$.  Bottom right:
Best 320 modes for optimising along the likelihood ridge axis.   Likelihood
contours are separated by 0.5 in natural log.\label{Manypar}}
\end{figure}

This latter case motivates an alternative strategy, which recognises that
the marginal error is dominated not by the curvature of the likelihood in
the parameter directions, but by the curvature along the principal axis
of the Hessian matrix with the smallest eigenvalue.    Figure 2 shows
how various strategies fare with a simultaneous estimation of the 
amplitude of clustering $\Delta$ and the redshift 
distortion parameter $\beta$, in a 
simulation of the PSCz galaxy redshift survey.   The top left panel shows
the likelihood surface for the full set of 508 modes considered for this
analysis (many more are used in the analysis of the real survey).  The modes 
used, and indeed the parameters involved, are not important for the
arguments here.   We see that the parameter estimates are highly correlated.
The second panel, top right, shows the single-parameter optimisation
of the first part of this paper.  The modes are optimised for $\beta$, 
and only the best 320 modes are used.  We see that the conditional error in 
the $\beta$ direction is not much worse than the full set, but the likelihood
declines slowly along the ridge, and the marginal errors on both $\beta$
and $\Delta$ have increased substantially.  In the panel bottom left, the
SVD procedure has been applied to the union of modes optimised for 
$\beta$ and $\Delta$,  keeping the best 320 modes.  The procedure does 
reasonably well, but in this case the error along the ridge has increased.
The bottom right graph shows the result of diagonalizing the Fisher 
matrix and optimising for the eigenvalue along the ridge.  We see excellent
behaviour for the best 320 modes, with almost no loss of information 
compared with the full set.   This illustrative example shows how
data compression may be achieved with good results by application of 
a combination of rigorous optimisation and a helping of common sense.

\section{Conclusions}

We have shown that single-parameter estimation by likelihood analysis 
can be made efficient in the sense that we can compress the 
original data set to make parameter estimation tractable, and it is 
optimal in the sense that there is no loss of information 
about the parameter we wish to estimate. Our eigenmodes are
generalised versions of the signal-to-noise eigenmodes, and 
are optimal for parameters entering the data covariance matrix in arbitrary 
ways.

As with all parameter estimation, this is a model-dependent
method in the sense that we need only to know the covariance matrix
of the data and the assumption of Gaussianity. However we have 
not had to introduce anything more than the standard assumptions
of likelihood analysis. The dependence on the initial choice of parameter
values is minimal, and can be reduced further by iteration.

For many-parameter estimation, we have shown the effects of two algorithms for
optimisation.  Optimising separately for several parameters by the 
single-parameter method, and trimming the resulting dataset via an SVD step
is successful in recovering the conditional likelihood errors.  For
correlated parameter estimates,  a promising technique appears to be to 
diagonalize the Fisher matrix and optimise for the single parameter 
along the likelihood ridge.

\end{document}